\documentclass[twocolumn,prl,aps,superscriptaddress,showpacs]{revtex4}
\usepackage{graphicx}

\preprint{DRAFT}

\begin{document}

\title{
Charge-Transfer Excitations in the Model Superconductor
HgBa$_2$CuO$_{\bf 4+\delta}$
}

\author{L. Lu}
\affiliation{Department of Applied Physics,
Stanford University, Stanford,
California 94305}

\author{X. Zhao}
\affiliation{Department of Physics, Jilin University, Changchun,
P.R.China}

\affiliation{Stanford
Synchrotron Radiation Laboratory, Stanford University, Stanford,
California 94309}

\author{G. Chabot-Couture}
\affiliation{Department of Applied Physics,
Stanford University, Stanford,
California 94305}

\author{J. N. Hancock}
\affiliation{Stanford Synchrotron Radiation Laboratory, Stanford
University, Stanford, California 94309}

\author{N. Kaneko}
\altaffiliation{Present address: National Metrology Institute of Japan, AIST,
Tsukuba 305-8568, Japan}

\affiliation{Stanford
Synchrotron Radiation Laboratory, Stanford University, Stanford,
California 94309}

\author{O. P. Vajk}
\altaffiliation{Present address: NIST Center for Neutron Research,
National Institute of Standards and Technology,
Gaithersburg, Maryland 20899, USA}

\affiliation{Department Physics,
Stanford University, Stanford,
California 94305}

\author{G. Yu}
\affiliation{Department Physics,
Stanford University, Stanford,
California 94305}

\author{S. Grenier}
\affiliation{Department of Physics and Astronomy, Rutgers
University, Piscataway, New Jersey 08854}
\affiliation{Department
of Physics, Brookhaven National Laboratory, Upton, New York 11973}

\author{Y. J. Kim}
\altaffiliation{Present address: Department of Physics, University of Toronto,
Toronto, Ontario M5S 1A7, Canada}

\affiliation{Department of Physics, Brookhaven National
Laboratory, Upton, New York 11973}

\author{D. Casa}
\affiliation{CMC-CAT, Advanced Photon Source, Argonne National
Laboratory, Argonne, Illinois 60439}

\author{T. Gog}
\affiliation{CMC-CAT, Advanced Photon Source, Argonne National
Laboratory, Argonne, Illinois 60439}

\author{M. Greven}
\affiliation{Department of Applied Physics,
Stanford University, Stanford,
California 94305}
\affiliation{Stanford
Synchrotron Radiation Laboratory, Stanford University, Stanford,
California 94309}

\date{\today}

\begin{abstract}
We report a Cu $K$-edge resonant inelastic x-ray scattering (RIXS)
study of charge-transfer excitations in the 2-8 eV range in the
structurally simple compound HgBa$_2$CuO$_{4+\delta}$ at optimal
doping ($T_{\rm c} = 96.5 $ K). The spectra exhibit a significant
dependence on the incident photon energy which we carefully utilize
to resolve a multiplet of weakly-dispersive ($ < 0.5$ eV)
electron-hole excitations, including a mode at 2 eV. The observation
of this 2 eV excitation suggests the existence of a remnant charge
transfer gap deep in the superconducting phase. Quite generally, our
data demonstrate the importance of exploring the incident photon
energy dependence of the RIXS cross section.

\end{abstract}


\maketitle

A pivotal challenge in the study of correlated electron systems is
to understand the nature of their electronic excitations. Compared
to our knowledge of magnetic and single quasi-particle excitations,
remarkably little is known about the hierarchy and momentum
dependence of elementary collective charge excitations in doped Mott
insulators such as the cuprate superconductors. The reason for this
limitation has been the lack of a suitable spectroscopic technique.
Optical spectroscopy  probes the dipole-allowed electron-hole pair
excitations, 
but is limited to zero momentum
transfer~\cite{uchida91}, while electron energy loss spectroscopy
(EELS) is surface sensitive and strongly affected by multiple
scattering effects at large momentum transfers \cite{wang96, fink01,
moskvin02, moskvin03}. The relatively new technique of resonant
inelastic x-ray scattering (RIXS), on the other hand, has the
potential both to probe such excitations with bulk sensitivity and
to yield momentum-resolved information throughout the Brillouin zone
\cite{kotani01}.

Previous RIXS studies focused predominantly on the undoped
insulators and revealed an excitation with an energy of $\sim 2$ eV
\cite{abbamonte99,hasan00,kim02}, consistent with
optical spectroscopy \cite{uchida91}. 
This charge-transfer (CT) excitation is associated with dipole transitions from the top of the
valence, or Zhang-Rice singlet (ZRS) band to the unoccupied
conduction band. In La$_2$CuO$_4$
\cite{kim02}, the excitation was found to disperse strongly ($\sim 1$ eV)
along [$\pi$,0] and was quickly suppressed along [$\pi$,$\pi$],
while in Ca$_2$CuO$_2$Cl$_2$ \cite{hasan00}  it was observed to
disperse by $\sim 0.7$ eV along [$\pi$,0], but by $\sim 1.6$ eV
along [$\pi$,$\pi$]. Finally, in Sr$_2$CuO$_2$Cl$_2$, it was found
to shift very weakly and to lose definition 
along [$2\pi,\pi$] \cite{abbamonte99}. In each case, RIXS also revealed additional
broad excitations at higher energy. The strongest such feature,
located at 6$-$7 eV, is commonly accepted to be a local molecular
orbital excitation, where an electron is excited from a bonding
to an antibonding state on a single CuO$_4$ plaquette
\cite{kim04_3}. An additional, weakly dispersing mode was observed
near 4 eV in La$_2$CuO$_4$~\cite{kim02}.
Two important issues that remain to be resolved are the apparent material dependent differences and the evolution with doping of the low-energy CT excitations.

We present RIXS measurements for optimally-doped
HgBa$_2$CuO$_{4+\delta}$ (Hg1201)  at high-symmetry positions along
$[\pi,0]$ and $[\pi,\pi]$. Hg1201 is a model superconductor due to
its high value of $T_c$ (highest among all single-layer cuprates),
simple tetragonal structure, relatively large spacing between
CuO$_2$ planes, and the absence of structural features
characteristic of other cuprates, such as superstructure
modulations and Cu-O chains \cite{eisaki04}. Through a careful analysis of the incident energy dependent RIXS spectra, we are able to identify a $\sim 2$ eV CT excitation and to determine an upper bound of 0.5 eV for its dispersion. The observation of this mode suggests the existence of a remnant CT gap in optimally-doped Hg1201. In addition, we are able to resolve weakly-dispersive excitations at $\sim 3$, 4, 5, 6, and 7 eV. We also
present new results for La$_2$CuO$_4$ which reveal that this
multiplet structure is present already in the undoped Mott
insulators. To account for this rich structure, we suggest that in addition to a
ZRS contribution, it is necessary to consider CT processes that
involve non-bonding O $2p$ bands with a small admixture of Cu $3d$
levels.

\begin{figure}
\includegraphics[height=6.2 cm,keepaspectratio=true] {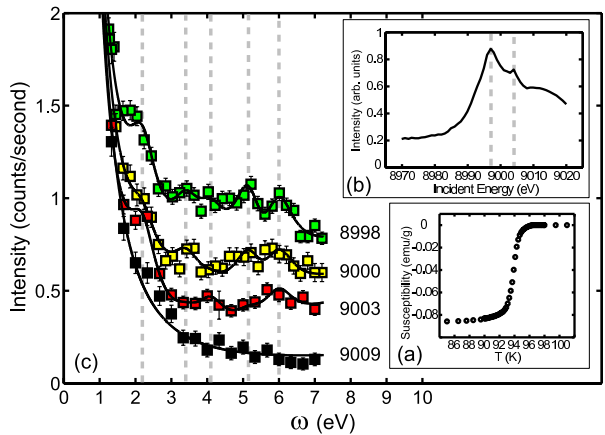}
\vspace{-8mm} 
\caption{\label{fig:hg1} Hg1201: (a) Magnetic susceptibility.
(b) X-ray absorption spectrum monitored by fluorescence yield. 
(c) Zone-center RIXS spectra at four incident energies, shifted
below 9009 eV by 0.2 counts/s relative to each other. Solid lines
are fits described in the text. Vertical lines indicate (b) the two intermediate
states where a resonant enhancement is observed in RIXS and
(c) distinct electron-hole excitations.
\vspace{-6mm} }
\end{figure}

A large ($\sim 20$ mm$^3$) Hg1201 single crystal was grown by the
flux method at Stanford University \cite{zhao05}. It was
heat-treated in an oxygen atmosphere to achieve optimal doping, and
magnetometry indicated $T_c = 96.5$ K (onset), as shown in Fig. 1(a).
Initial characterization at the Stanford Synchrotron Radiation
Laboratory, with 12 keV x-rays at beam line 7-2, revealed a mosaic
of 0.04$^{\circ}$ (FWHM). 
RIXS was performed on the
undulator beam line 9IDB at the Advanced Photon Source, with
incident photon energy ($E_{\rm i}$) set to values near the Cu
$K$-edge using a Si(333) double monochromator. A spherical, diced
Ge(733) analyzer was used, and the overall energy resolution was 300
meV (FWHM). The experiment was performed around the weak (2,0,1) and (3,0,0)
reflections. The incident photon polarization was perpendicular to
the vertical scattering plane and within the CuO$_2$ planes.
Figure 1(b) shows the x-ray absorption spectrum for our crystal.
The main edge (8998 eV) and satellite peak (9003 eV) correspond to the two distinct  $1s\rightarrow4p_\sigma$ transitions responsible for the resonant 
enhancement present in the RIXS spectra. 
The La$_2$CuO$_4$ crystal had a N\'eel temperature of
$T_{\rm N} = 320$ K \cite{vajk02} and was measured with polarization
perpendicular to the CuO$_2$ planes, consistent with previous work
\cite{kim02}. All measurements were carried out at room temperature.
In this Letter, we specify the in-plane component of the reduced
wave vector ${\bf q}$ and use units with lattice constant $a\equiv1$
and $\hbar \equiv 1 $. The energy transfer is defined as $\omega =
E_{\rm i} - E_{\rm f}$.

Figure \ref{fig:hg1}(c) shows Hg1201 zone-center spectra at four values of
$E_{\rm i}$. For $E_{\rm i} = 8998$ eV, we observe a peak at $\omega
= 2.2$ eV together with two less intense, but clearly resolvable
features at 5.2 eV and 6.0 eV, and a spectral weight enhancement at
intermediate energy transfers. By gradually tuning $E_{\rm i}$ to
higher energies, two additional features, centered at 3.4 eV and 4.1
eV, sequentially stand out when compared to the featureless
``background" spectrum at $E_{\rm i} = 9009$ eV. We will refer to
these features as ``2 eV," ``3 eV," ``4 eV," ``5 eV," and ``6 eV,"
according to their approximate  peak positions.
We emphasize that, when viewed as a second-order process, the RIXS cross section, as in the case of resonant Raman scattering \cite{blumberg97}, depends on both 
$\omega$ and $E_{\rm i}$ \cite{kotani01}.

\begin{figure}
\includegraphics[height=8.7 cm,keepaspectratio=true] {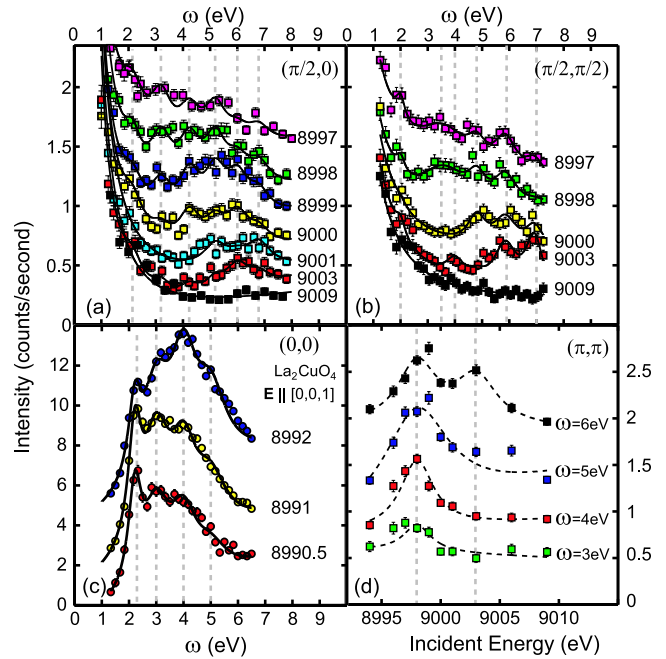}
\vspace{-3mm} 
\caption{\label{fig:hg2} Hg1201 RIXS spectra at (a) ($\pi/2,0$) and
(b) ($\pi/2,\pi/2$). The data are shifted below 9009 eV by 0.2
counts/s relative to each other. The solid lines are fits described
in the text. (c) La$_2$CuO$_4$ zone-center sprectra; 
8991 and 8992 eV data are shifted vertically. (d) Hg1201 $E_{\rm
i}$-dependence at ($\pi$,$\pi$) for several energy transfers. }
\vspace{-6mm}
\end{figure}

The relative strength of the excitations also is a function of wave
vector. Figure \ref{fig:hg2} contains results for (a) ($\pi/2$,0)
and (b) ($\pi$/2,$\pi$/2). The 2 eV mode rapidly decreases in
strength toward the zone boundary,
similar to previous results for La$_2$CuO$_4$ \cite{kim02}.
On the other hand, the higher-energy modes gain in intensity toward
the zone boundary and the multiplet structure becomes more
pronounced. Apart from the 2 eV and 3 eV features, which are not
discernible at ($\pi$,0), and ($\pi$,$\pi$), respectively, the
multiplet structure in the $2$-$6$ eV range is similar to the zone
center result [Fig. 1(c)]. We focus on the five lowest-lying
features, but note that some spectra exhibit additional 
features at $\sim7$ eV and $\sim9$ eV (not shown).

Although RIXS measures the electron-hole pair dynamics rather than
the properties of a single hole quasiparticle, the multiplet
structure resembles valence band photoemission results for
Sr$_2$CuO$_2$Cl$_2$, for which separate bands with rather different
mixing between Cu $3d$ and O $2p$ orbitals have been associated with
the significant intensity variations with momentum and
energy \cite{pothuizen97,hayn99}.

\begin{figure}
\includegraphics[height=6 cm,keepaspectratio=true] {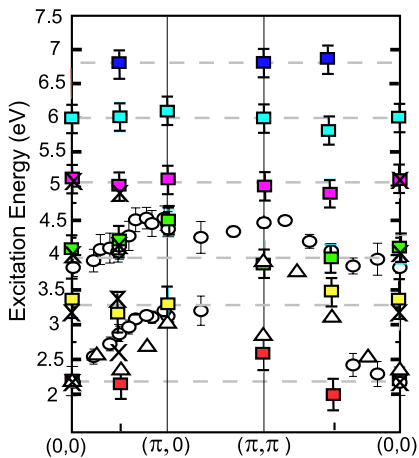}
\vspace{-3mm} 
\caption{\label{fig:hg3} Dispersion of CT excitations for
optimally-doped Hg1201 (squares),
compared with previous results below 6 eV for
Ca$_2$CuO$_2$Cl$_2$
(triangles)~\cite{hasan00,zahid_comment}
and La$_2$CuO$_4$ (circles)~\cite{kim02}, 
and new data for La$_2$CuO$_4$ (crosses).}
\vspace{-6mm}
\end{figure}

A reasonable estimate of the peak positions can be obtained already
by visual inspection of the data. To arrive at a more precise
estimate, at each wavevector we first fit the $E_{\rm i}$ = 9009 eV
data to a Lorentzian centered at $\omega =0$, and then used this as
``background" in a {\it simultaneous} fit of all the other spectra
to a set of Lorentzians with $E_{\rm i}$-independent peak positions
and $E_{\rm i}$-dependent amplitudes. The description of some of the
spectra required an additional linear contribution.
The peaks are broader than our resolution of 0.3 eV, and
in the fits of Figs. 1 and 2 the peak widths were fixed at a value of 0.8 eV 
\cite{taguchi97,kotani99,kotani00} based on their apparent widths.

As can be seen from Fig. \ref{fig:hg3}, the difference between the
excitation energies at (0,0) and the zone boundary is relatively
small for all modes, within 0.5 eV. Rather than observing a single
highly dispersive mode below 4 eV, as found previously for the Mott
insulators La$_2$CuO$_4$ and Ca$_2$CuO$_2$Cl$_2$, we discern two
separate and much less dispersive branches \cite{zahid_comment}.
Because of the significant $E_{\rm i}$-dependence observed here
for Hg1201, it appears possible that two weakly dispersive
low-energy modes already are present in the undoped Mott insulators, 
which could unify the seemingly disparate behavior of different
compounds  \cite{abbamonte99,hasan00,kim02}. To test this possibility, we
performed a new measurement for La$_2$CuO$_4$ [at (0,0), shown in 
Fig. 2 (c), and also at  ($\pi/2,0$)], and a reanalysis of the zone-center result of
Ref.~\cite{kim02}
along the lines discussed above, which suggests that this is indeed
the case. The data indicate four features below 6 eV at energies 
very similar to those for
Hg1201 (see Fig. 3). It
therefore seems very likely that the strong dispersion observed in
Refs. \cite{hasan00,kim02} stems from an effective admixture of the
two lowest-lying modes. Nevertheless, from our result at ($\pi/2,0$)
it appears that the 2 eV mode in the Mott insulator has a larger
dispersion than in superconducting Hg1201, consistent with
recent theory 
\cite{tsutsui03} and experiment
\cite{hasan04}.

Recent RIXS studies of doped cuprates
suggest an interesting doping dependence of the spectral features
\cite{kim04,hasan04,ishii04}. A continuum of intensity was found to
emerge below 2 eV with increasing doping for
La$_{2-x}$Sr$_x$CuO$_4$ ~\cite{hasan04,kim04}, and the lowest-energy 
CT excitation (at $\sim$ 2 eV) is completely suppressed for
$x=0.17$~\cite{kim04}. Due to the relatively strong ``elastic tail"
we were not able to probe the region well below 2 eV in Hg1201 with the present
energy resolution.
However, in contrast to the observation for La$_{2-x}$Sr$_x$CuO$_4$,
which appears to be consistent with optical spectroscopy \cite{uchida91},
we clearly resolve a well-defined low-energy feature at $\sim 2$ eV.
It rapidly decreases in strength away from the zone
center, but can still be traced up to the zone boundary, with a net dispersion 
no greater than 0.5 eV. 
We note that  it is unlikely to be a $d-d$ type excitations. The latter lie below 2 eV
\cite{kuiper98,ghiringhelli04} and are expected to be much weaker 
at the $K$-edge than at the $L$ \cite{ghiringhelli04} and
$M$ edges \cite{kuiper98}. Furthermore, at the Cu $K$-edge, one might expect
such excitations  to be significant at the pre-edge ($E_{\rm i}$ = 8985 eV, 
$1s-3d$ transition), but we did not observe a ``2eV" feature at this incident energy.
Therefore, optimally doped Hg1201, the single-layer cuprate with the highest $T_{\rm c}$,
exhibits a remnant CT gap.
It would be very valuable to extend initial 
results for the optical
conductivity of  Hg1201 \cite{homes04}.

The most significant feature for $E_{\rm i} = 9003$ eV is the non-dispersing 
6 eV mode, which is very likely the (local) molecular orbital excitation from 
the bonding (predominantly 3$d^{9}$) to the antibonding state (predominantly
3$d^{10}\underline{L}$), consistent with a systematic study among cuprates of the 
bond-length dependence  of the energy of this feature 
($d_{Cu-O}$ =  1.9396 \AA~at $T = 300$ K for our crystal)~\cite{kim04_3}. 
As further indication of its local character, we note that this excitation resonates at both 
$E_{\rm i}$ = 8998 eV and 9003 eV, as can be seen from Fig. 2(d). These
two resonance energies are the final states in the x-ray absorption spectra of Fig. 1(b),
and they correspond to different intermediate states
in the RIXS process, one which is well-screened (8998 eV; predominantly
$1\underline{s}3d^{10}\underline{L}4p$) and the other
poorly-screened (9003 eV; predominantly $1\underline{s}3d^{9}4p$).
The double-resonance behavior has been
captured using the Anderson impurity model
of a single CuO$_4$ plaquette \cite{kotani99,kotani00}, thus enforcing the conclusion that these excitations are local.

With the exception of the 2 eV feature, the Hg1201 spectra for
$\omega < 6$ eV exhibit a different resonant behavior as they are
much stronger at $E_{\rm i} = 8998$ eV than at 9003 eV [Fig. 2(d)].
This suggests that the corresponding excitations have a more significant overlap
with the well-screened (3$d^{10}\underline{L}$) than with the
poorly-screened (3$d^9$) intermediate states. 
In this scenario, the resultant screening can have both local and
nonlocal contributions, as firstÊ
suggested in the context of Cu $2p$ core-level photoemission 
~\cite{koitzsch02,veenendaal93,boske98,karlsson99,tranquada91}. 
In nonlocal screening, a hole residing on the
surrounding oxygen ligand is Coulomb-repelled by the $1s$ core hole
on the central Cu site and delocalizes to a neighboring CuO$_4$
plaquette. The delocalized hole can either form a ZRS with a
neighboring Cu, or move to non-bonding O $2p$ orbitals ($2p_{\pi}$,
$2p_{\sigma}$ and $2p_{z}$). Due to the planar polarization in
our experiment, CT transitions to purely $2p_{z}$ orbitals are
suppressed.
Therefore, there are three nonlocal modes that may
contribute to the observed spectra. 
In local screening, on the
other hand, the single-Cu site Anderson impurity model calculations \cite{kotani99,kotani00}
reveal that the final state has nearly pure O $2p$ character due to
the strong core hole potential on the same plaquette. Unlike for the
nonlocal modes, the ligand hole does not migrate to a neighboring
CuO$_4$ plaquette. For in-plane polarization, there should be two
local modes that correspond to CT transitions to the two local
in-plane O $2p$ orbitals ($2p_{\sigma}, 2p_{\pi}$). Consequently,
we arrive at five candidate modes for the four observed features below
6 eV. Due to the intrinsic broadening and the relatively high
density of modes, we might not be able to discern all of them.

Initial EELS measurements of Sr$_2$CuO$_2$Cl$_2$
indicated a single CT gap excitation of $\sim2$ eV with a large
dispersion of 1.5 eV along [$\pi,\pi$] ~\cite{wang96}. However, a
more recent high-resolution ($115$ meV) study revealed a multiplet
structure in this Mott insulator
~\cite{fink01,moskvin02}, analogous to our present observations.
By extending theoretical models to include a complete set of Cu $3d$ and O $2p$
states, the multiplet structure 
was found to be consistent with a prediction of multiple charge-transfer
type excitations, originating either locally or nonlocally.
Recent models of the Cu $K$-edge RIXS processes, 
based on a multi-band Hubbard model approach, show that at half-filling 
a total of four transitions appear below 6 eV \cite{nomura04,nomura05,markiewicz05}. 
In principle, the addition of holes permits additional transitions 
into the ZRS band which may not be easily discernible.
At optimal (hole) doping, the theoretical prediction for the RIXS spectra is remarkably
similar to the result at half-filling for a reasonably large choice of Cu onsite repulsion
\cite{markiewicz05}. 
This is consistent with our experimental findings for Hg1201 and La$_2$CuO$_4$.

In conclusion, we have measured electron-hole pair excitations in the
2-8 eV range along [$\pi,0$] and [$\pi,\pi$] in optimally-doped
HgBa$_2$CuO$_{4+\delta}$ using resonant
inelastic x-ray scattering. By carefully utilizing the incident
photon-energy dependence of the cross section, we discern a
multiplet of excitations, and we establish an upper bound of 0.5 eV
for the dispersion of all excitations. Our observation of an excitation at
$\sim 2$ eV in this very-high-$T_{\rm c}$ material is in surprising contrast with
previous findings for (La,Sr)$_2$CuO$_4$.
One possible reason for this difference may be the presence of significant
quenched disorder in close proximity to the Cu-O sheets in the latter material
\cite{eisaki04}.
While superconducting HgBa$_2$CuO$_{4+\delta}$ is expected to have significant spectral weight 
at lower energies, our finding suggests the existence of a remnant charge-transfer gap. 
Now that sizable HgBa$_2$CuO$_{4+\delta}$ crystals have become available,
we plan to complement the present work on this model superconductor
using other experimental techniques in order to fully elucidate materials specific differences
among the cuprate superconductors.
Our partial reinvestigation of La$_2$CuO$_4$ revealed the multiplet structure
already present in the Mott insulating parent compounds, and it indicates that
the dispersion of the 2 eV mode is smaller than previous estimates.
Importantly, our work establishes that the incident energy dependence
of RIXS spectra  of correlated materials
is critical information that needs to be carefully considered in
future experiments and theory, and it calls for the use of a multi-band
theoretical approach.

We would like to thank S. Larochelle, K.M. Shen, T.H. Geballe, J.P. Hill,
R.S. Markiewicz, Z.X. Shen, T. Tohyama, and J. Zaanen for helpful discussions.
The work at Stanford University was supported
by the DOE under Contracts No.
DE-FG03-99ER45773 and No. DE-AC03-76SF00515.


\begin{thebibliography}{999}
\bibitem{zahid_comment} The ($\pi,\pi$) data of Ref. \cite{hasan00} are consistent with the presence of two modes. In Fig. 3, we plot the observed peak positions rather than the center
of gravity of the features observed in Ref. \cite{hasan00}.
\end{thebibliography}
\end{document}